\documentclass[11pt]{article}
\usepackage[dvips,usenames]{color}
\usepackage[dvips]{pstcol} 

\usepackage{graphicx}

\def \be{\begin{displaymath}}
\def \ee{\end{displaymath}}              

\def \ben{ \begin{equation} }
\def \een{ \end{equation}   }            

\def \bea{\begin{eqnarray*}}             
\def \eea{\end{eqnarray*}}

\def \bean{\begin{eqnarray}}             
\def \eean{\end{eqnarray}}

\def \nn{\nonumber}

\def \Pr{\mbox{Pr}}

\def \span{ \mbox{span} }
\def \tr{ \mbox{{\rm tr}}}

\def \eps{\varepsilon}

\def \ccc#1{ {\cal #1} }               
\def \b#1{ \mathbf{#1} }               

\def \Ref#1{(\ref{#1})}

\def \dag{\;\!\!^\dagger}

\def \tenpow#1{^{\otimes #1}}                  

\def \invb#1 { \frac{1}{#1} }
\def \bra#1{\langle #1 |}                      
\def \ket#1{| #1 \rangle }                     

\def \sp#1{ \langle #1 \rangle }               

\def \cav#1{ [ #1 ] }                          
\def \scav#1{ \left[ #1 \right] }              

\def \fr#1#2{ \frac{#1}{#2} }

\def \norm#1{\parallel \! #1 \! \parallel}

\def \tn#1{\norm{#1}_{tr}}           

\def \fn#1{\norm{#1}_{F}}            

\def \scrbox#1{{\scriptsize\mbox{#1}}}

\begin{document}
\input epsf

\title{Approximate quantum error correction, random codes, and quantum
  channel capacity}
\author{Rochus Klesse\footnote{Email address: rk@thp.uni-koeln.de}\\[0.5cm]
\centerline{ {\sl Universit\"at zu K\"oln, Institut f\"ur Theoretische
    Physik, Z\"ulpicher Str. 77,}}\\
\centerline{ {\sl  D-50937 K\"oln, Germany}} }
\date{April 24, 2007}
\maketitle
\begin{abstract}
We work out a theory of approximate quantum error correction
that allows us to derive a general lower bound for the entanglement fidelity
of a quantum code. The lower bound is given in terms of Kraus
operators of the quantum noise. This result is then used to analyze
the average error correcting performance of codes that are randomly
drawn from unitarily invariant code ensembles. Our results confirm
that random codes of sufficiently large block size are highly suitable
for quantum error correction. Moreover, employing a lemma of 
Bennett, Shor, Smolin, and Thapliyal, we prove that random coding
attains information rates of the regularized coherent information. 
\end{abstract}

\section{Introduction}\label{sec-introduction}
Physical processing, transmission, and storage of quantum information 
unavoidably suffers from decohering interactions with the environment.
The insight that the resulting errors can, in
principle, be corrected has been a major breakthrough in the field of
quantum information theory \cite{Sho95,Ste96}. A theory of quantum
error correction (QEC) rapidly evolved \cite{EM96,BDSW96,KL97} and
eventually led to the concept of quantum 
fault tolerance \cite{Pre98}, which, in fact, put large-scale quantum computation
back in the realms of possibility.  
Quantum error correction stands in close relation to the information
capacity of a noisy quantum channel and the quantum coding theorem
\cite{Llo97,BNS98,BKN00}. 

In this paper we elaborate a theory of approximate 
QEC. We obtain a general and easily computable lower bound for
the entanglement fidelity of a noisy channel $\ccc N$ that is attainable 
when the information is encoded in a given error correcting code. 
The bound is expressed in terms of Kraus operators of
$\ccc N$, and the projection on the code space. (Sec.\ \ref{sec-S-I}) 

We employ this theory to analyze the average error correcting
performance of codes that are chosen at random from certain code 
ensembles. For the unitarily invariant ensemble of all
$K$-dimensional code spaces we find a surprisingly simple lower bound
for the averaged code entanglement fidelity. Its deviation from unity 
is determined by $\sqrt{K N} \fn{ \ccc N(\pi_Q)
}$, where $N$ is the number of Kraus operators in an operator-sum
representation of the noise $\ccc N$ under consideration, $\pi_Q$ is the
homogeneously distributed state
of the system $Q$ on which $\ccc N$ is operating, and $\fn{A}$ denotes
the Frobenius 
norm $\sqrt{\tr A\dag A}$ of an operator $A$.
We derive this result by reverting to 
random matrix theory. 
For the special case of unital noise the lower bound immediately
reveals that randomly chosen codes 
attain with high probability the quantum Hamming bound \cite{EM96}
(Sec.\ \ref{sec-random}). 

Our next issue is the extension of the foregoing considerations to  
the case of noise operations that do not conserve the trace. We find it
useful to understand them as the result of a selective process 
and therefore define fidelities and coherent information in this
situation slightly different from the standard definitions in
literature (see, e.g.,\ \cite{NCSB98}) 
(Sec.\ \ref{sec-general-noise}). 

One motivation why we extend our theory to trace-decreasing operations
becomes apparent in the last section. Here we show that with the aid
of a recent lemma of 
Bennett, Shor, Smolin, and Thapliyal (BSST) \cite{BSST02,Hol02} our
results allow a relatively simple proof of 
the direct coding theorem. Our proof follows ideas of Shor \cite{Sho02}   
and Lloyd \cite{Llo97} by showing that QEC based on random code spaces 
attains rates of the regularized coherent information. 
The proof is therefore quite different from Devetak's one \cite{Dev05},
which is based on a correspondence of classical private information and 
quantum information (Sec.\ \ref{sec-lower}).

After having clarified some conventions and notations, we will start
in Sec.\ \ref{sec-approximate} with a brief introduction to 
QEC and quantum channel capacity. 
The remaining sections are organized as laid out above. 

\subsection{Conventions and notations}

 We denote a general mixed state by $\rho$, a general pure state
 by $\psi$, and add subscripts to indicate the system. 
 For instance, $\psi_{QR}$ means a
 pure state of the joint system of $Q$ and $R$. 

 We will use the trace norm $\tn{A} = \tr \sqrt{A\dag A}$, and the
 Frobenius (Hilbert-Schmidt) norm $\fn{A} = \sqrt{ \tr{ A\dag A} }$
 for a linear operator $A$. 
 The two-state fidelity is here defined as 
 $F(\rho_1, \rho_2) = \tn{ \sqrt{ \rho_1} \sqrt{ \rho_2} }^2 $.

\section{Quantum error correction  and quantum
  capacity}\label{sec-approximate}   
Throughout the paper, we consider a quantum system $Q$ that is supposed to
store or transmit quantum information. We denote the Hilbert space of
$Q$ by $H_Q$ and its finite dimension by $M$.  
In addition to a possible internal unitary dynamics,
$Q$ is subjected to external noise during storage or
transmission. Let the effect of both be described by a completely
positive, trace-preserving mapping $\ccc N$  that maps an initial density
operator $\rho$ to a final density operator $\rho' = \ccc N(\rho)$ 
\cite{Kra83,NC00}. 
We call $\ccc N$ either a noise operation or, synonymously, a
noisy channel. $\ccc N$ can be always represented in an
operator sum
\be
 \ccc N(\rho) =  \sum_{i=1}^N A_i \rho A_i\dag\:,
\ee
where the (non-unique) Kraus operators $A_1, \dots , A_N$
are linear operators on $H_Q$. They satisfy the completeness relation
$\sum_i A_i\dag A_i = \b 1_{Q}$.  

\subsection{Quantum error
  correction}\label{subsec-quantum-error-correction} 

In general, a QEC scheme for the noise $\ccc N$ on $Q$ is based on 
a quantum error correcting code $C$, which, by definition, is a certain
linear subspace $C$ of $H_Q$. Let $K$ be the dimension of $C$, and let
$P$ be the projection on $C$. We call a state $\rho$ a 
state in $C$ or a code state (of $C$)  if the support of
$\rho$ is a subset of $C$. 
If the code $C$ is suitably chosen, one may find a recovery
operation $\ccc R$ that exactly recovers all code state 
from corruption by $\ccc N$, i.e., for all code states $\rho$
of $C$, $\ccc R \circ \ccc N(\rho) = \rho$. 

Finding an optimal code $C$ for the correction of some
given noise $\ccc N$ is a difficult task. 
The code $C$ should be of course as large as possible, but at the same time
the encoding in $C$ must also be sufficiently redundant such that
errors caused by $\ccc N$ can still be identified and corrected.
In practice, the code may also satisfy additional technical
constraints. 
Somewhat simpler than this problem but nevertheless instructive is the
following related one: Given the noise operation $\ccc N$, what can be
gained by the use of a certain quantum code $C$? Here, theory does
provide definite answers in the form of necessary and sufficient
conditions for the feasibility of quantum error correction.

First, there are quite elementary necessary and sufficient conditions for 
exact QEC \cite{EM96,BDSW96,KL97,NC00}:

{\em Exact recovery of all code states is possible if and only if 
for all $i,j$ the operators $P A_j\dag A_i P$ are
proportional to $P$, 
\ben\label{recovery_condition}
P A_j\dag A_i P = \fr{1}{K} ( \tr P A_j\dag A_i P )\: P\:.
\een}

For explicitly given Kraus operators $A_i$ it is usually no problem to 
check these conditions. If they are satisfied, it is also possible to 
explicitly construct the Kraus operators for the recovery operation
$\ccc R$. 
Things become more complicated when the conditions are violated.  
In this case, it can become quite difficult to foresee whether the violation 
is serious, and therefore error correction virtually impossible, or
whether the violation is harmless and code states are still essentially
correctable up to some small deviations. An early approach to this
problem has been given in \cite{LNCY97}.

An alternative condition for QEC
can be formulated in terms of coherent information \cite{Llo97,SN96}.
The coherent information  $I(\rho, \ccc N) $ of
a state $\rho$ with respect to the noise $\ccc N$
is defined by  
\be
I(\rho, \ccc N) = S( \ccc N(\rho)) - S( \ccc I_R \otimes \ccc N (\psi_{RQ} ))\:,
\ee
where $S(\varrho) = - \tr\: \varrho \log_2 \varrho$ is the von Neumann
entropy, $\psi_{RQ}$ is a purification of  
$\rho$, and $\ccc I_R$ is the identity operation on the ancilla system
$R$. The last term, $S( \ccc I_R \otimes \ccc N(\psi_{RQ} ))$, is the  
entropy exchange $S_e(\rho,\ccc N)$ of $\rho$ with respect to $\ccc N$
\cite{Sch96}.
The coherent information obeys an important inequality \cite{SN96}: 
For any two operations  $\ccc E_1$ and $\ccc E_2$ 
\ben\label{data-processing-inequality}
S(\rho) \ge I(\rho, \ccc E_1 ) \ge I(\rho, \ccc E_2 \circ \ccc E_1 )\:.
\een
Moreover, equality in the first inequality holds if and only if 
the action of $\ccc E_1$ on $\rho$ can be completely reversed,  
meaning that there exists an $\ccc R$ such that 
$\ccc I_R \otimes (\ccc R \circ \ccc E_1) (\psi_{RQ}) =
\psi_{RQ}$, for any purification $\psi_{RQ}$ of $\rho$.  
This leads to the following necessary and sufficient
condition for error correction \cite{SN96}:  

{\em
Exact recovery of all code states is possible 
if and only if for a state $\rho_C$ with
$\mbox{{\rm supp}}\:(\rho_C)= C$ 
\ben\label{Schumacher-Nielsen}
S( \rho_C) = I(\rho_C, \ccc N)\:.
\een}

Schumacher and Westmoreland \cite{SW02} have shown that this
condition is robust against small
perturbations, i.e., if it is only 
approximately satisfied, then errors can still be approximately
corrected. 
Their central result is a lower bound for the entanglement
fidelity \cite{Sch96} of an arbitrary state $\rho$ under the noise $\ccc N$ and a
subsequent recovery operation $\ccc R$. 
It is proven that for given $\rho$ and 
$\ccc N$ there exists an $\ccc R$ such that
\ben\label{schumacher-westmoreland}
F_e(\rho, \ccc R \circ \ccc N) \: \ge \: 1 - 2 \sqrt{ S(\rho) - I(\rho, \ccc
  N) }\:.
\een
To elaborate on this, let us discuss
entanglement fidelity and its relevance for our purposes.

\subsection{Entanglement fidelity}\label{subsection-entanglement_fidelity}
 The entanglement fidelity $F_e(\rho,\ccc E)$ of a state  
$\rho$ under an operation $\ccc E$ on $Q$ is defined by
\be
F_e(\rho, \ccc E) := \sp{ \psi_{RQ} | \ccc I_R \otimes \ccc E (\psi_{RQ}) |
  \psi_{RQ}} \:, 
\ee
where $\psi_{RQ}$ is any purification of $\rho$. That this is
independent of the chosen purification can be seen 
from the representation in terms of Kraus operators of $\ccc N$,
$F_e(\rho, \ccc E) = \sum_{i=1}^N |\tr \rho A_i|^2$ \cite{Sch96}. 
Especially interesting is
the entanglement fidelity of the homogeneously distributed code state
$\pi_C = P/K$.  The reason is that $F_e(\pi_C, \ccc E)$ is a lower bound
of the {code-averaged channel fidelity} $F_{av}(C, \ccc E)$ 
(Appendix \ref{subapp-fe_le_fav}; cf. \cite{HHH99,Nie02}). 
Moreover, it can be shown that when $F_e(\pi_C, \ccc E)$ is close to unity, 
$C$ must have a large subcode $\tilde C \subset C$ 
with a similar high {minimum fidelity} $F_{min}(\tilde C, \ccc E)$
(Appendix \ref{subsec-subcodes}).
The entanglement fidelity $F_e(\pi_C, \ccc E)$ is therefore a convenient
figure of merit that characterizes the distortion of states in $C$
under the operation $\ccc E$. 

In order to capture the suitability of a code $C$ for QEC without referring 
to a certain recovery operation we introduce 
\ben\label{code-entanglement-fidelity}
F_e(C, \ccc N) := \max_{\ccc R} F_e(\pi_C, \ccc R \circ \ccc N)\:,
\een
the entanglement fidelity of the code $C$ under noise $\ccc N$.
By relation \Ref{schumacher-westmoreland} it is then clear that 
\ben\label{Fe-inequality}
F_e(C, \ccc N) \ge 1 - 2 \sqrt{ S(\pi_C) - I(\pi_C,
  \ccc N) }\:.
\een
This shows that for small $S(\pi_C) - I(\pi_C,  \ccc N) \ll 1$ the
code entanglement fidelity is close to unity and thus approximate QEC
is possible. 

Building on ideas of Schumacher and Westmoreland's proof of relation
\Ref{schumacher-westmoreland}, here we will derive an alternative
lower bound for the code entanglement fidelity $F_e(C, \ccc N)$ that
is explicitly given in terms of the Kraus operators of $\ccc N$ (cf.\ relation 
\Ref{final-Fe-inequality} in Sec.~\ref{sec-S-I}).
However, before we start, let us briefly point out that the
code entanglement fidelity \Ref{code-entanglement-fidelity} can also be
used to conveniently define quantum capacity of a noisy channel.

\subsection{Quantum capacity of a noisy
  channel}\label{subsec-quantum_capacity} 
We consider the following scheme of information transmission from
Alice (sender) to Bob (receiver) by means of the channel $\ccc N$
\cite{BNS98}: 
Alice is allowed to encode quantum
information in blocks of $n$ identical copies of $Q$, with the block
size $n$ and the encoding operation $\ccc E_n$ at her disposal. 
Sending the block to Bob, each individual system $Q$ is independently 
disturbed by the noise operation $\ccc N$, i.e., the whole block
$Q^n$ is subjected to $\ccc N\tenpow n$. Bob is allowed to perform any
decoding operation $\ccc R_n$ in order to restore the message which
Alice originally sent. The maximum amount of quantum information,
measured in units of qubits, that can be reliably transmitted per 
channel use in such a scheme defines the quantum capacity $Q(\ccc N)$ of
the noisy channel $\ccc N$ \cite{BNS98}. 

Precise mathematical definitions of the quantum capacity can be given
in many ways \cite{KW04}. Here we use one that fits in the present
context of approximate QEC and the code entanglement fidelity. 

It has been shown that restricting the encoding operation $\ccc E_n$ to 
isometric embeddings into $H_Q\tenpow n$ has no effect on the
capacity \cite{BKN00}. $\ccc E_n$ is thus sufficiently described by the
subspace $C_n$ 
of $H_Q\tenpow n$ whose code states represent the encoded information.
Viewing $C_n$ as an error correcting code, Bob is able to reconstruct
Alice's message within a precision that is given by the code
entanglement fidelity  
$F_e(C_n, \ccc N\tenpow n)$. 
We follow the standard definitions and call $R$ an achievable rate
of $\ccc N$ if there is a sequence of code spaces $C_n \subset
H_Q\tenpow n$, $n=1,2,\dots$, such that  
\ben\label{achievable_rate}
\lim_{n\to \infty} \sup \fr{ \log_2 \dim C_n }{n} = R\:,\quad 
\mbox{and}
\quad
\lim_{n\to \infty} F_e(C_n, \ccc N\tenpow n) = 1\:.
\een
The quantum capacity $Q(\ccc N)$ is the supremum of all achievable rates
$R$ of $\ccc N$.   

The quantum coding theorem for noisy channels
\cite{Llo97,BNS98,BKN00} states that 
the quantum capacity $Q(\ccc N)$ of a channel $\ccc N$ equals the
regularized coherent information 
\ben\label{def-rci}
I_r(\ccc N) = \lim_{n \to \infty} \fr{1}{n} \max_{\rho } I(\rho, \ccc N\tenpow n)\:.
\een
$I_r(\ccc N)$ has long been known an upper bound for $Q(\ccc N)$, which is
the content of the converse coding theorem \cite{BNS98,BKN00}. 
The direct coding theorem, stating that $I_r(\ccc N)$ is actually
attainable, has been strictly proven only recently by Devetak
\cite{Dev05}. His proof utilizes the 
correspondence of private classical information and quantum information. 
More direct proofs in the spirit of Shannon's ideas on random coding 
\cite{SW49}
have been earlier outlined by Shor \cite{Sho02} and Lloyd \cite{Llo97}.
In the last section, we will employ our theory to provide a strict
proof along these lines.

\section{Lower bound for the code entanglement
  fidelity}\label{sec-S-I} 
In this section we derive a lower bound for the code entanglement 
fidelity $F_e(C, \ccc N)$ in terms of Kraus operators $A_1,\dots, A_N$
of $\ccc N$ and the projection $P$ on the $K$-dimensional code $C$.  
We will show that 
\ben\label{final-Fe-inequality}
F_e(C, \ccc N)  \ge 1 - \tn D\:,
\een
where
\ben\label{Dop-explicit}
D 
= \fr{1}{K} \sum_{ij=1}^N
\left(P A_i\dag A_j P - \fr{1}{K}(\tr P A_i\dag A_j P) P \right)
\otimes \ket i \bra j \:, 
\een
is an operator on $C \otimes H_E$, with $H_E$ being an ancilla
Hilbert-space spanned by orthonormal vectors $\ket 1, \dots, \ket N$. 

The coefficients of $D$ precisely correspond to the conditions
\Ref{recovery_condition} for exact error correction. If these are
fulfilled the operator $D$ vanishes and inequality
\Ref{final-Fe-inequality} also predicts perfect error 
correction. In this sense, the lower bound \Ref{final-Fe-inequality}
can be considered as a generalization of the elementary conditions 
\Ref{recovery_condition} to the case of approximate QEC. 
It is worth mentioning that the lower bound does not depend on the 
chosen set of Kraus operators $A_1, \dots, A_N$ for $\ccc
N$. Equivalent sets are related by a unitary transformation
\cite{NC00} which in Eq.\ \Ref{Dop-explicit} amounts merely to a unitary
basis change, and therefore leaves $\tn{D}$ invariant. 

To prove relation \Ref{final-Fe-inequality} we describe $\ccc N$
as a unitary $U_{QE}$ on $Q$ and an environment $E$, followed by
a partial trace over $E$ \cite{Kra83,NC00}. That is, for a general state $\rho_Q$ 
\be
\ccc N(\rho_Q) = \tr_E\: U_{QE} \:\rho_Q \otimes \psi_E \:
U_{QE}\dag\:,
\ee
where $\psi_E$ is some fixed initial state of $E$. Further, let
$\psi_{RQ}$ be a purification of $\rho_Q$, let $\rho_R = \tr_Q\:
\psi_{RQ}$, and let a pure state $\psi_{RQE}'$ on $RQE$ be defined by
\be
\psi_{RQE}' = ( \b 1_R \otimes U_{QE} )
\: \psi_{RQ}\otimes \psi_E 
\: ( \b 1_R \otimes
U_{QE}\dag )\:.
\ee
$\psi_{RQE}'$ purifies its partial states 
\bean
\rho_Q' &=& \tr_{RE}\: \psi_{RQE}'\:, \qquad
\rho_E' = \tr_{RQ}\: \psi_{RQE}'\:, \nn\\
\rho_{RE}' &=& \tr_{Q}\: \psi_{RQE}'\:, \qquad
\rho_{RQ}' = \tr_{E}\: \psi_{RQE}'\:. \label{partial_states}
\eean
Following ideas that has been utilized in \cite{SW02} and \cite{ADHW06} we 
show that there exists a recovery  operation $\ccc R$ on $Q$ such that 
\ben\label{decoupling}
F_e(\rho_Q, \ccc R \circ \ccc N) \ge 1- \tn{ \rho_{RE}' - \rho_{R}
  \otimes \rho_E' }\:.  
\een
The idea is to find in the vicinity of 
the actual final state $\psi_{RQE}'$ (or an extension $\psi'$ of
it) a state $\tilde \psi$ from  which $\psi_{RQ}$ can be perfectly recovered
by an operation $\ccc R$ on $Q$. The distance between $\psi'$ and
$\tilde \psi$ will then determine a lower bound for the entanglement 
fidelity $F_e(\rho_Q, \ccc R \circ \ccc N)$. 

To this end, we consider the product state 
$ \rho_R \otimes \rho_E' $ with its obvious purification 
\be
\tilde \psi := \psi_{RQ} \otimes \psi_{RQE}' 
\ee
on the joint system $RQSE$, where $S$ denotes a copy of $RQ$. 
We extend $\psi_{RQE}'$ to a pure state $\psi'$ on $RQSE$ 
by some pure state $\psi_{S}$ of $S$
(i.e. tracing out $S$ or $RQE$ yields $\psi_{RQE}'$ or
$\psi_{S}$, respectively).  
According to Uhlmann's theorem \cite{Uhl76,Joz94,NC00}
there is a unitary $U_{QS}$ on $QS$ such that 
\ben\label{uhlmann}
|\sp{ \tilde \psi |\: U_{QS}\: \psi'}|^2 = F(\rho_R \otimes
\rho_E', \rho_{RE}')\:. 
\een
Then, for a recovery operation $\ccc R$ on Q defined by 
\be
\ccc R(\rho_Q) := \tr_{S} \:U_{QS} \:\rho_Q \otimes \psi_S\: U_{QS}\dag
\ee
we find 
\be \ccc I_R \otimes \ccc R (\rho_{RQ}') = \tr_{SE}\: U_{QS}\:
\psi'\: U_{QS}\dag\:,
\ee 
which by the monotonicity of the fidelity under
partial trace \cite{NC00} and $\psi_{RQ} = \tr_{SE}\: \tilde \psi$ yields
\be
F_e(\rho_Q, \ccc R \circ
\ccc N) \equiv  
F(\psi_{RQ}, \ccc I_R \otimes \ccc R (\rho_{RQ}')\: )
\ge  |\sp{ \tilde \psi | \: U_{QS}\: \psi'}|^2 \:.
\ee
With Eq.\ \Ref{uhlmann} and the general relation $F(\rho,\sigma) \ge 1
-  \tn{ \rho-\sigma}$ \cite{NC00} this proves relation
\Ref{decoupling}.

Now, we become more specific and chose for given Kraus operators 
$A_1,\dots, A_N$ of $\ccc N$ its representing unitary $U_{QE}$ such that 
\ben\label{unitary_UQE}
U_{QE} \ket{ \psi_Q } \ket{1} = \sum_{i=1}^N A_i \ket{ \psi_Q}
\ket{i}\:,
\een
where $\ket 1 \equiv \ket{\psi_E}, \ket{2}, \dots, \ket{N}$ are
orthonormal vectors in  $H_E$. Further, let $\rho_Q = \pi_C \equiv
P/K$ with purification 
\ben\label{psi_RQ}
\ket{\psi_{RQ}} = \fr{1}{\sqrt{K}} \sum_{l=1}^K \ket{c_l^R}
\ket{c_l^Q}\:, 
\een
where the orthonormal vectors $\ket{ c_1^R}, \dots \ket{ c_K^R}$ 
and $\ket{ c_1^Q}, \dots \ket{ c_K^Q}$ span $H_R$ and $C$,
respectively. For this setting, we obtain  
\bean
\rho_{RE}' &=& \fr{1}{K} \sum_{ij=1}^N \sum_{l,m=1}^K
\tr_Q( A_i \ket{c_l^Q}\bra{c_m^Q} A_j\dag)\:
\ket{c_l^R}\bra{c_m^R}\otimes \ket{i}\bra{j}\:, \label{rho_RE'}\\
\rho_R \otimes \rho_E' &=& \sum_{ij=1}^N \tr_Q( A_i \pi_C A_j\dag)\:
\rho_R \otimes \ket{i}\bra{j}\:.  \label{rho_R'_times_rho_E'}
\eean
Things become more convenient if we isometrically map both states with 
an isometry defined by 
\be 
J:
\sum_{ij,lm}  \alpha_{ij,lm} \ket{c^R_l}\bra{c^R_m} \otimes \ket i
\bra j
\quad \mapsto \quad
\sum_{ij,lm}  \alpha_{ij,lm}^* \ket{c^Q_l}\bra{c^Q_m} \otimes \ket i
\bra j
\ee
to 
\bea
X &:= & J( \rho_{RE}') \qquad = \: \fr{1}{K} \sum_{ij=1}^N P A_i\dag A_j
P \: \otimes \ket{i}\bra{j}\\
Y &:=& J( \rho_R \otimes \rho_E')\: = \:\fr{1}{K} \sum_{ij=1}^N
\fr{1}{K} \tr( P A_i\dag A_j P)  \: P \: \otimes \ket{i}\bra{j}\:.
\eea 
Hence, $\tn{ \rho_{RE}' - \rho_{R}\otimes \rho_E'} = \tn{X-Y}$,  
which with relation \Ref{decoupling} leads us to  
\be
F_e(\pi_C, \ccc R \circ \ccc N) \ge 1 - \tn{X-Y}\:. 
\ee 
Since the left-hand side is a lower bound of the code entanglement
fidelity $ F_e(C, \ccc N)$, and $X-Y = D$, this finally proves
relation \Ref{final-Fe-inequality}.

\section{Random quantum codes}\label{sec-random}
Random codes play an important role in classical as well as in 
quantum information theory. In this section we will analyze the
average error correcting performance of random codes by means of
the lower bound \Ref{final-Fe-inequality} for the entanglement fidelity
of the codes. 
We consider the same setting as before: a quantum information storing
system $Q$ with $M$-dimensional Hilbert space $H_Q$ that is exposed to 
noise $\ccc N$ with a set of Kraus operators $A_1, \dots A_N$. 

\subsection{Ensemble averaged code fidelity}\label{subsec-ensemble_averaged}
Let ${E_K}$ be an ensemble of $K$-dimensional codes in $H_Q$
with an ensemble average  $\cav{A}$ defined for code dependent
variables $A= A(C)$.
We are interested in the ensemble averaged code entanglement fidelity 
$ \cav{F_e(C, \ccc N)}$. By inequality \Ref{final-Fe-inequality},  
\ben\label{fin}
\cav{ F_e(C, \ccc N) } \ge 1 - \scav{ \tn D }\:,
\een
where $D$ is the code dependent operator Eq.\ \Ref{Dop-explicit}.

In many cases, averaging the trace norm of $D$ would be quite a difficult
undertaking. We therefore prefer to estimate $\scav{ \tn D}$ 
by the more convenient average of the squared Frobenius norm, $\cav{ \fn{ D }^2} = \cav{\tr
  D\dag D }$:
Trace norm and Frobenius norm of $D$ with domain $C \otimes H_E$
of dimension $d=KN$ satisfy  
\be 
\tn D  \le \sqrt{d}  \fn D \:.
\ee 
We remark that this inequality is a good estimate only if the
eigenvalues of $D$ are of similar  
magnitude. Using this estimate and employing Jensen's inequality 
\cite{jensens_inequality} we obtain
\ben\label{trace-frobenius}
\cav{ \tn D } \le \sqrt{d} \cav{ \fn D}
=  \sqrt{d} \scav{ \sqrt{ \fn{ D }^2} } 
\le \sqrt{ d \scav{  \fn{ D }^2} }\:, 
\een
and so 
\ben\label{fe-final}
\cav{ F_e(C,\ccc N) } 
\:\ge \:
1 - \sqrt{ KN \scav{ \fn{ D }^2 }}\:.
\een
In the next subsection we will evaluate this lower bound for  
unitarily invariant code ensembles. 

\subsection{Unitarily invariant code ensembles}\label{subsec-unitarily_invariant}
Let $U_K$ be the unitarily invariant code ensemble that consists
of all $K$-dimensional codes in $H_Q$, furnished with 
the unitarily invariant ensemble average
\be 
\cav{ A(C)}_{U_K} := \int_{\b U(H_Q)} d\mu(U)\: A( U C_0)\:,
\ee
where $C_0 \subset H_Q $ is some fixed code space of dimension $K$,
and $\mu$ is the (normalized) Haar measure on $\b U(H_Q)$, the group
of all unitaries on $H_Q$. Later on we will  
also consider an analogously defined ensemble $U_K(V)$ that consists
of $K$-dimensional codes in some subspace $V$ of $H_Q$.

Our task is to calculate $\cav{ \fn{ D }^2 }_{U_K}$.
By the explicit representation Eq.~\Ref{Dop-explicit} of operator
$D$ we immediately find 
\be
\fn{ D }^2\: =  \tr\: D\dag D = \fr{1}{K^2} \sum_{ij=1}^N \tr( P W_{ij}\dag
P W_{ij} ) - \fr{1}{K} |\tr PW_{ij} |^2\:,
\ee
where the operators $W_{ij}$ are
\be
W_{ij} = A_i\dag A_j\:.
\ee
The ensemble average of $\fn{ D }^2$ 
can be conveniently calculated if we introduce a Hermitian form 
\ben\label{b-definition} 
b(V,W) := \scav{ \tr( P V\dag P W ) - \fr{1}{K} \tr( P V\dag) \: \tr(P
  W) }_{U_K}\:,
\een
such that 
\ben\label{b-average}
\scav{ \fn{ D }^2 }_{U_K} = \fr{1}{K^2} \sum_{ij} b(W_{ij}, W_{ij})\:.
\een
We recall that $P$ is the projection on the $K$-dimensional code
space that is chosen with unitarily invariant probability from
the ensemble ${U_K}$. 
By Eq.\ \Ref{b-definition} it is therefore clear that 
$b(V,W)$ is a unitarian invariant on $H_Q$, i.e., for  any $U \in \b U(H_Q)$
\be
b(U V U\dag, U W U \dag) = b(V, W)\:.
\ee
This places us in a position to utilize the general theory of group invariants
by Weyl \cite{Wey46,How92}:
In the present situation it means 
that $b(V,W)$ must be a linear combination of the 
two fundamental unitarily-invariant Hermitian forms
$ \tr\: V\dag W$ and $ \tr\:V\dag  \tr\: W$,
\ben\label{b-linear}
b(V,W) =  \alpha\: \tr\: V\dag W \: + \: \beta\: \tr\:V\dag  \tr\: W\:.
\een
To determine the coefficients $\alpha$ and $\beta$ we derive 
two linear independent equations by equating Eqs.~\Ref{b-definition}
and \Ref{b-linear} for two special choices of the operators $V$ and
$W$. For $V=W=\b 1_{H_Q}$ we obtain as a first equation,
\ben\label{I}
    \alpha\: M + \beta \: M^2 = 0\:.
\een 
Next, we set $V=W=P_1$, where $P_1$ is the projection on 
an one-dimensional space spanned by some unit vector $\ket{\psi} \in H_Q$.
From Eq.~\Ref{b-definition} we immediately  find
\be
b(P_1, P_1) = \left(1-\fr{1}{K} \right) \scav{ | \sp{\psi | P | \psi}
  |^2 }_{U_K}\:.
\ee
Reverting to results from random matrix theory, we obtain in
Appendix \ref{app-rmt}  
$\scav{ | \sp{\psi | P | \psi}|^2}_{U_K} = (K^2+K)/(M^2+M)$ 
(which for large $K$ and $M$ is close to the naive estimate $\scav{ |
  \sp{\psi | P | \psi}|^2}_{U_K} \approx  \scav{\sp{\psi | 
  P | \psi} }_{U_K}^2 = K^2/M^2$). Thus,
\be
b(P_1, P_1) = \fr{K^2-1}{M^2+M}\:.
\ee
With $ b(P_1, P_1) = \alpha  +  \beta $ from
Eq.~\Ref{b-linear} this yields the second equation,
\ben\label{II}
\alpha\: + \: \beta \: =  \fr{K^2 -1}{M^2 + M}\:.
\een
Solving Eq.s~\Ref{I} and \Ref{II} for $\alpha$ and $\beta$, and
inserting the solution into Eq.\ \Ref{b-linear} produces
\be
b(V,W) = \fr{K^2-1}{M^2-1} \left( \tr\:V\dag W \: 
- \: \fr{1}{M} \tr\:V\dag \: \tr\:W \right)\:,
\ee
and, by Eq.~\Ref{b-average}, 
\ben\label{exact-average}
\scav{ \fn{ D }^2 }_{U_K} =\fr{1-1/K^2}{M^2-1} \sum_{ij} 
\left(\tr\:W_{ij}\dag W_{ij}  
- \fr{1}{M} |\tr\:W_{ij}|^2  \right)\:.
\een
In general, not much is given away if instead of this exact result we
use an upper bound for $\scav{ \fn{ D }^2}_{U_K}$ that we obtain by using
$(1-1/K^2)/(M^2-1) \le 1/M^2$ and by omitting the negative terms
$-|\tr\:W_{ij}|^2/M $ in the sum. Then
\be
\scav{ \fn{ D }^2 }_{U_K} \: \le \: \fr{1}{M^2} \sum_{ij} 
\tr\:W_{ij}\dag W_{ij} 
= \tr(\sum_j A_j \fr{\b 1}{M} A_j\dag \sum_i A_i \fr{\b 1}{M} A_i\dag
)\:,
\ee
where we cyclically permuted operators under the trace to obtain the
last equality. We realize that the argument of the trace is simply $\ccc
N(\pi_Q)^2$,  with $\pi_Q = \b 1_Q/M$ being the homogeneously
distributed density operator on $H_Q$. This yields the rather simple
upper bound 
\ben\label{final-D-upperbound}
\scav{ \fn{ D }^2 }_{U_K} \: \le \:   \fn{ \ccc N( \pi_Q)}^2\:.
\een 
By relation \Ref{fe-final} this means 
\ben\label{Fe-lowerbound}
\cav{ F_e(C, \ccc N) }_{U_K} \ge 1 -  \sqrt{KN} \fn{ \ccc N(\pi_Q)} \:.
\een

Before discussing this result let us 
generalize it to the unitarily invariant ensemble $U_K(V)$ of
$K$-dimensional codes in a subspace $V\subset H_Q$ ($\dim V \ge
K$). Here the average is given  
by
\be
\cav{ A(C) }_{U_K(V)} := \int_{\b U(V)} d\mu_V(U)\: A( U C_0)\:,
\ee
where $\mu_V$ is the normalized Haar measure on the group $\b
U(V)$ of unitaries on the subspace $V$. 
Up to the fact that now the role of $H_Q$ is taken over by the linear
space $V$ nothing has changed compared to the situation before. 
Hence, the derivation given above for the ensemble $U_K$ applies to
the ensemble $U_K(V)$ as well, 
showing that
\ben\label{D-upperbound-V}
\cav{ \fn{ D }^2 }_{U_K(V)} \: \le \: \fn{ \ccc N(\pi_V) }^2\:,
\een
and consequently, 
\ben\label{Fe-lowerbound-V}
\cav{ F_e(C, \ccc N) }_{U_K(V)} \: \ge \: 1 -  \sqrt{KN} \fn{ \ccc N(\pi_V)
}\:,
\een
where $\pi_V = \Pi_V/ \dim V$.  

\subsection{Discussion}\label{subsec-discussionII}
It is instructive to discuss the just obtained lower bounds for the case 
of unital noise, which by definition leaves the homogeneously
distributed state $\pi_Q$ invariant, $\ccc N(\pi_Q) = \pi_Q$. A unital
operation is for instance  
the process where arbitrary unitary operations
$U_1, \dots, U_N$ are applied to the system $Q$ with probabilities
$p_1, \dots, p_N$. For unital noise 
$\fn{ \ccc N(\pi_Q)}^2 = \fn{ \pi_Q}^2 = \tr (\pi_Q^2) = 1/M$. 
Hence, by the lower bound \Ref{Fe-lowerbound}, 
\be
\cav{ F_e(C, \ccc N)}_{U_K} \: \ge \: 1 - \sqrt{\fr{KN}{M}} \:.
\ee
This means that on almost all codes $C$ of the ensemble $U_K$
the unital noise $\ccc N$ can be almost perfectly corrected, provided
that 
\be 
KN  \ll M \:.
\ee
Recalling that $K$ is the code dimension, $N$ is the number of
Kraus operators in an operator-sum representation of $\ccc N$, and $M$ is the 
dimension of $H_Q$, we recover that randomly chosen codes attain
the quantum Hamming bound \cite{EM96}. 

The requirement $K \ll M/N$ suggests that $\log_2 M - \log_2 N$ is
a lower bound of the capacity $Q(\ccc N)$, what we will now formally
derive.
To this end, we consider the $n$-fold replicated noise $\ccc N\tenpow n$,
and study 
the averaged  entanglement fidelity of the code ensemble 
$U_{K_n}$, where we chose the code dimension to be $K_n = \lfloor
2^{nR} \rfloor$ for some positive $R$.
$\ccc N\tenpow n$ operates on states in $H_Q\tenpow n$ and has $N^n$
operation elements. With $\ccc N$ also $\ccc N\tenpow n$ is
unital, thus $\fn{ \ccc N\tenpow n( \rho_{Q_n})}^2 = M^{-n}$, and 
by Eq.\ \Ref{Fe-lowerbound}
\be
\cav{ F_e(C, \ccc N\tenpow n)}_{U_{K_n}} \: \ge \: 1 - \left(
  \fr{2^{R}\: N}{M}   \right)^{n/2}\:. 
\ee
In the limit $n\to \infty$ the right hand side converges to unity if 
$R < \log_2 M - \log_2 N$. Since $\lim_{n\to \infty} \fr{1}{n} \log_2 K_n
= R$ this implies that all rates below $\log_2 M-\log_2 N$ are
achievable and so, by the definition of quantum capacity in
\ref{subsec-quantum_capacity},   
\be
Q(\ccc N) \ge \log_2 M - \log_2 N\:.
\ee 
We note that since $\ccc N$ is unital $\log_2 M = S(\pi_Q) = S(\ccc
N(\pi_Q))$. Now, if we could identify the second term, $\log_2 N$, with the
entropy exchange $S_e(\pi_Q, \ccc N)$ we would obtain that the lower bound
$\log_2 M - \log_2 N$ is just the coherent information $I(\pi_Q, \ccc
N)$, in accordance to the capacity formula. 
However, this is the case only for a special kind of 
unital operations. $\ccc N$ must have a Kraus representation with 
operation elements $A_1, \dots, A_N$ such that $\tr A_j\dag A_i = 0$
for $i\neq j$, and $\fr{1}{M}\tr A_i\dag A_i = const. = 1/N$. Then 
by Schumacher's relation indeed 
\be
S_e(\pi_Q, \ccc N) = S \left( \{ \tr A_i \pi_Q
  A_j\dag \}_{i,j=1,\dots N} \right) 
= S( \b 1_N / N) = \log_2 N\:.
\ee
The first condition is actually no restriction, since a
nondiagonal representation $B_1, \dots, B_N$ with $\tr B_i\dag B_j
\neq 0$ can always be unitarily transformed to a diagonal one
(cf.\ footnote \ref{foot}). The second condition demands that, roughly speaking,
different kinds of errors appear with equal probability. In the end, 
this ensures that by the estimation $\tn{ D } \le \sqrt{KN} \fn{ D}$ not
much is lost and therefore the lower bound \Ref{Fe-lowerbound} is good.

To recapitulate, for unital noise $\ccc N$  the lower bounds for the
ensemble averaged code fidelities immediately make evident that 
the quantum Hamming bound is attainable by random codes. Moreover, 
if the noise $\ccc N$ satisfies the condition of equally probable errors as 
specified above we can establish 
\ben\label{coherent-information-lower-bound}
Q_{\ccc N} \ge I(\pi_Q, \ccc N)\:.
\een

\section{Error correction in selective noise}\label{sec-general-noise}
The hitherto presented analysis is restricted to trace-preserving
noise operations. Here we will extend the considerations of the preceding
sections to the case of trace-decreasing noise, which we find to be
convenient in later use. First, we define channel fidelity and
entanglement fidelity for a trace-decreasing channel.
Within this definitions we will then generalize the lower bound
\Ref{final-Fe-inequality} and the result \Ref{Fe-lowerbound-V} on the
ensemble averaged code fidelity.  

\subsection{Fidelities for trace-decreasing channels}\label{subsec-selective-channels}
For a (possibly) trace-decreasing operation $\ccc N$ on a system $Q$ we
define the channel fidelity with respect to a state $\rho_Q$ as 
\ben\label{selective-channel-fidelity}
F_{ch}(\rho_Q, \ccc N) := \tr \ccc N(\rho_Q) \: F(\rho_Q, \fr{ \ccc N(\rho_Q)}{\tr
  \ccc N(\rho_Q)} )\:, 
\een
where $ F(\rho, \sigma)$ is the usual two-state fidelity. 
The definition deviates from the standard one by a factor 
$\tr \ccc N(\rho_Q)$. This makes sense, when one interprets a trace-decreasing $\ccc
N$ as a selective operation that selects
individual elements of the initial ensemble $\rho_Q$ with probability 
$\tr \ccc N(\rho_Q)$ \cite{Kra83}. Consequently, in order that
$F_{ch}(\rho_Q, \ccc N 
)$ is close to unity not only the selected final state $\ccc
N(\rho_Q)/ \tr \ccc N(\rho_Q)$ must be close to $\rho_Q$, but also the
selection probability must be close to unity.   

We define the entanglement fidelity of $\ccc N$ with respect to
$\rho_Q$ as 
\ben\label{selective-entanglement-fidelity}
F_e(\rho_Q, \ccc N) := F_{ch}(\psi_{RQ}, \ccc I_R \otimes \ccc N)
= \sp{ \psi_{RQ} | (\ccc I_R \otimes {\ccc N})
  (\psi_{RQ}) |\psi_{RQ}} \:, 
\een
where $\psi_{RQ}$ purifies $\rho_Q$.  
Note that if $\ccc N$ is trace-decreasing also its extension $\ccc I_R
\otimes {\ccc N}$ is trace-decreasing, in which case $F_{ch}$ means the
just defined fidelity \Ref{selective-channel-fidelity}.
Repeating the arguments of Schumacher \cite{Sch96}, it is not
difficult to see that also the 
entanglement fidelity of a trace-decreasing $\ccc N$ can be expressed
by its Kraus operators $A_1, \dots,
A_N$ of $\ccc N$ by the usual formula
\ben\label{sel-Fe}
F_e(\rho_Q, {\ccc N}) = \sum_{i=1}^{N} |\tr \rho_Q A_i |^2\:.
\een
A simple but important consequence of this relation is the
following: Let for a subset $\tilde N \subset \{1, \dots, N\}$ 
a quantum operation $\tilde \ccc N$ be defined by
\be
\tilde \ccc N(\rho_Q) := \sum_{i \in \tilde N} A_i \rho_Q A_i\dag\:, 
\ee
which we will call a reduction of the operation $\ccc N$. 
Then by Eq.\ \Ref{sel-Fe},
\be
F_e(\rho, \ccc N) \ge F_e(\rho, \tilde{ \ccc N})\:.
\ee
Further, since for any operation $\ccc R$ on $Q$ clearly $\ccc R \circ
\tilde{\ccc N}$ is a reduction of $\ccc R \circ \ccc N$, we conclude that
for any code $C$
\ben\label{selection-lowers-fidelity}
F_e(C, \ccc N) \ge F_e(C, \tilde{\ccc N})\:,
\een
where the code entanglement fidelity for a trace-decreasing ${\ccc N}$ 
is defined as for trace-preserving noise by 
$ F_e(C, {\ccc N}) := \max_{\ccc R} F_e(\pi_C, \ccc R \circ {\ccc N}) $.

\subsection{Lower bound for code entanglement fidelity} 
Let $\ccc N$ be a noise operation on $Q$ that can be represented by
Kraus operators $A_1, \dots, A_N$.  
The entanglement fidelity of a $K$-dimensional code $C$ satisfies
\ben\label{selective-Fe-inequality}
F_e(C, \ccc N) \: \ge \: \tr\ccc N(\pi_C) \: - \tn{ D }\:, 
\een
where $\pi_C = P/K$ is the homogeneously distributed code state,
and the operator $D$ is defined exactly as in Eq.\ \Ref{Dop-explicit}. 

This relation generalizes the lower bound \Ref{final-Fe-inequality} to
the case of a trace-decreasing operation $\ccc N$. Its proof given in 
Appendix \ref{app-selective-fidelity} is almost identical to the one of 
Eq.\ \Ref{final-Fe-inequality} in Sec.~\ref{sec-S-I}. 

\subsection{Unitarily invariant code ensembles}
We consider the ensemble $U_K(V)$ of all $K$-dimensional codes in a
subspace $V$ of $H_Q$ which we introduced 
in Sec.\ \ref{subsec-unitarily_invariant}.
According to the lower bound \Ref{selective-Fe-inequality}, the
averaged code entanglement fidelity under a (possibly trace-decreasing)
noise $\ccc N$ with Kraus operators $A_1, \dots , A_N$ satisfies
\be
\scav{ F_e(C,\ccc N)}_{U_K(V)} \: \ge \: 
\scav{ \tr \ccc N(\pi_C) }_{U_K(V)} 
\: - \:
\scav{ \tn{ D } }_{U_K(V)}
\:, 
\ee
where $D$ is given by Eq.\ \Ref{Dop-explicit}. As shown in
Sec.\ \ref{sec-random}, 
\be
\scav{ \tn{ D } }_{U_K(V)} 
\: \le \: 
\sqrt{ K N \scav{ \fn{ D }^2 } }_{U_K(V)} 
\: \le \: 
\sqrt{K N} \fn{ \ccc N(\pi_V)
}\:, 
\ee
where $\pi_V = \Pi_V/\dim V$ (cf. Eq.s \Ref{trace-frobenius} and
\Ref{D-upperbound-V}). Furthermore, we will show below that 
\ben\label{tr-tildeN}
\cav{ \tr {\ccc N}(\pi_C) }_{U_K(V)} = \tr {\ccc N}(\pi_V)\:, 
\een
and thus obtain 
\ben\label{fe-tildeN-V}
\cav{F_e(C, {\ccc N})}_{U_K(V)} \: \ge \:
\tr {\ccc N}(\pi_V) 
\: - \:
\sqrt{K N} \fn{ \ccc N(\pi_V)
}\:.
\een

We show Eq.\ \Ref{tr-tildeN} by again referring to unitarian invariants:
Let a linear form $a$ on the set of all linear operators on $H_Q$ be
defined by 
\be
W \mapsto a(W) := \fr{1}{K} \cav{ \tr P W P }_{U_K(V)}\:,
\ee
where, as always, $P=\Pi_C$. Since the codes $C$ are subspaces of 
$V$ it is clear that $a(W) = a( \Pi_V W \Pi_V)$. 
Further, the  unitarian invariance of the code ensemble entails $a(W) = a(U W
U\dag)$ for all unitary transformations $U$ on $H_Q$ with $U(V) = V$. 
It follows that $a$ must be proportional to the 
fundamental invariant linear form on $V$, $W\mapsto \tr(\Pi_V W)$.
From $a(\Pi_V)=1$ we can then deduce that $a(W) = \tr(\pi_V W)$. To
conclude the proof of Eq.\ \Ref{tr-tildeN} we 
note that 
\be
\cav{ \tr {\ccc N}(\pi_C) }_{U_K(V)}
= \sum_{i=1}^{ N} \fr{1}{K}\cav{ \tr A_i P A_i\dag}_{U_K(V)}
= \sum_{i=1}^{ N} a(A_i\dag A_i) 
= \tr \sum_{i=1}^{ N}A_i
\pi_V A_i\dag = \tr {\ccc N}(\pi_V)\:.
\ee

\section{Lower bounds for the quantum capacity}\label{sec-lower}
In this section we will prove that the quantum capacity $Q(\ccc N)$ of
a general trace-preserving channel $\ccc N$ satisfies
\ben\label{rho_V-lowerbound}
Q(\ccc N) \ge I(\pi_V, \ccc N)\:,
\een
where $\pi_V$ is the homogeneously distributed density on an arbitrary
subspace $V$ of the system's Hilbert space $H_Q$.  
We will then use the lemma of BSST in order to establish
the regularized coherent information $I_r(\ccc N)$ (cf.\ \Ref{def-rci})
as a lower bound of $Q(\ccc N)$. 

We first prove inequality \Ref{rho_V-lowerbound} for the case
$V= H_Q$ or $\pi_V =\pi_Q$. A strategy of proof
becomes evident when we look back at Sec.\ \ref{subsec-discussionII},
where we showed 
$Q(\ccc N) \ge I(\pi_Q,\ccc N)$
under the conditions of
\begin{description}

\item (i)  equally probable errors, and
\item (ii) unitality: $\ccc N(\pi_Q) = \pi_Q$.

\end{description}
For general noise $\ccc N$ these two requirements are certainly not
fulfilled, not even approximately. However, since our concern is the 
channel capacity of $\ccc N$ we are free to consider the $n$-times replicated
channel $\ccc N \tenpow n$.
For large $n$ it is possible to arrange for the conditions (i) and
(ii) in an approximate sense by, as it will turn out, only minor
modifications of the operation $\ccc N \tenpow n$. Following Shor
\cite{Sho02}, we  
\begin{description}
\item (a) reduce the operation $\ccc N\tenpow n$ to an operation $\ccc N_n$
  that consists only of the {\em typical} Kraus operators of $\ccc
  N_n$ (cf.\ Sec.\ \ref{subsec-typical}). 
\end{description}
Thereafter we
\begin{description}

\item (b) project 
          on the typical subspace $T_n$ of $\ccc N(\pi_Q)$ in
          $H_Q\tenpow n$ (cf.\ Sec.\ \ref{subsec-projection}).

\end{description} 
The purpose of reduction (a) is to approximately establishes a situation of
equally probable errors (i). The second step allows to restrict the
output Hilbert space of $\ccc N_n$ to the typical subspace $T_n$, on
which the density $\ccc N_n(\pi_{Q_n})$ is approximately homogeneously
distributed. This establishes a situation similar to (ii).
After having proven Eq.\ \Ref{rho_V-lowerbound} for $V=H_Q$ in
Sec.\ \ref{subsubsec-rhoQ}, we will argue in Sec.\ \ref{subsubsec-rhoV}
that its generalization is trivially obtained by restricting the
original input Hilbert space $H_Q$ of $\ccc N$ to a subspace $V \subset
H_Q$.  
Finally, in Sec.\ \ref{subsubsec-rho} we use the lemma of BSST
in order to show that $Q(\ccc N) \ge I_r(\ccc N)$.

\subsection{Reduction of the noise $\ccc N \tenpow n$}\label{subsec-reduction}

Both, typical Kraus operators and typical subspaces are 
defined on the basis of typical sequences (see, e.g., \cite{NC00}). We
briefly recall their definition and state two basic facts that are
important for  
our purposes. 

\subsubsection{Typical sequences}\label{subsubsec-typical_sequences}
Let $X_1,\: X_2, \: \dots, X_n$ be a sequence of independent random
variables that assume values $A_1, \dots, A_N$ with probabilities
$p_1,\: p_2, \dots, p_N$. We denote the probability distribution by
$\ccc A$. Its Shannon entropy is 
$
H(\ccc A) = - \sum_{i=1}^N p_i \log_2\: p_i\:. 
$
Let $\eps$ be some positive number. A sequence $\b A = A_{j_1}, A_{j_2},
\dots, A_{j_n}$ is defined to be $\eps$-typical if its probability 
of appearance 
$
p_{\b A} = p_{j_1} p_{j_2} \dots p_{j_n}
$
satisfies 
\be
    2^{- n( H(\ccc A) + \eps)} \: \le \: p_{\b A} \: \le \: 2^{-n( H(\ccc
      A) - \eps)}\:.
\ee
Below we will make use of the following two facts:
\begin{enumerate}

\item the number of all $\eps$ typical sequences $N_{\eps,n}$ is less
  than $2^{n(H(\ccc A) +\eps)}$,

\item the probability $P_{\eps,n}$ that a random sequence of length $n$ 
  is $\eps$-typical satisfies \\ $ 1- P_{\eps,n} \: \le \: 2 e^{-n \psi(\eps)}\:,
$
where $\psi(\eps)$ is a positive number independent of $n$.
\end{enumerate}
Proofs can be found in Appendix \ref{app-typical}.

\subsubsection{Restriction to typical Kraus operators}\label{subsec-typical}
Let a trace-preserving noise $\ccc N$ on $Q$ be represented by 
Kraus operators $A_1, \dots, A_N$.
Without loss of generality we can assume that the $A_i$ are
diagonal in the sense that $\tr A_j\dag A_i =0$ for $i\neq j$.
\footnote{\label{foot} For arbitrary operation elements $B_1,
  \dots, B_N$ of $\ccc N$ let an $N\times N$ matrix $H$ be defined by
$
H_{ij} := \tr B_i\dag B_j\:.
$
Since $H=H\dag$, there is a unitary matrix $U$ such that $U H U\dag$
is diagonal. Because of the unitary freedom in the operator-sum
representation \cite{NC00},
the operators $A_m:= \sum_j U\dag_{jm} B_j$ equivalently represent $\ccc
N$. It is readily verified that $\tr A_l\dag A_m = 0$ for $l\neq m$.}
We define {\em the probability $p_i$ of the Kraus operator $A_i$} 
as
\ben\label{operation-probs}
p_i := \fr{1}{M} \tr A_i\dag A_i\:, 
\een
and we denote the corresponding probability distribution by $\ccc
A_{\ccc  N}$.  
The definition makes sense, because the $p_i$ are positive and,
as a consequence of the trace preservation of $\ccc N$, sum up to unity.

The $n$-times replicated noise $\ccc N\tenpow n$ can be represented by 
$N^n$ Kraus operators
\be
A_{j_1} \otimes A_{j_2} \otimes \dots \otimes A_{j_n} \equiv A_{\b j}  \:,
\ee
where $j_\nu = 1, \dots, N$ and $\b j = (j_1,j_2,\dots,j_n)$.
By the diagonality of the operators $A_i$ of $\ccc N$ also the operators
$A_{\b j}$ of $\ccc N\tenpow n$ are diagonal, and the probability $p_{\b j}$ 
of the element $A_{\b j}$ appears to be the product of the
probabilities $p_{j_\nu}$ of its constituent elements $A_{j_\nu}$,
\be
p_{\b j} = \fr{1}{M^n} \tr A_{\b j}\dag A_{\b j} 
 =  \fr{1}{M^n} 
\tr (A_{j_1}\dag A_{j_1}) \: \dots\:
\tr(A_{j_n}\dag A_{j_n})  
 =  p_{j_1} \dots p_{j_n}\:.
\ee
In other words, the Kraus operators $A_{\b j}$ of $\ccc N\tenpow n$ 
are sequences of length $n$ in which symbols $A_i$ of an
alphabet $A_1, \dots, A_N$ appear according to the distribution $\ccc
A_{\ccc N}$. 
Hence we are in the domain of classical random sequences and
can employ the notions of Sec.\ \ref{subsubsec-typical_sequences}
to define the $\eps$-typical operation $\ccc N_{\eps,n}$ of 
$\ccc N\tenpow n$ by 
\be
\rho \mapsto \ccc N_{\eps, n}(\rho)\: :=  \sum_{ A_{\b j}
  \scrbox{ $\eps$-typical} }
  A_{\b j} \: \rho A_{\b j}\;\!\! \dag\:,
\ee
i.e., $\ccc N_{\eps,n}$ consists only of the $\eps$-typical Kraus
operators of $\ccc N$. In general, this strongly reduces the number of 
Kraus operators from $N^n$ to 
\be 
N_{\eps,n} \: \le \: 2^{n( H(\ccc A_{\ccc N}) + \eps)}
\ee
(cf.\ Sec.\ \ref{subsubsec-typical_sequences}, property 1.).
It is time to remark that $H(\ccc A_{\ccc N})$ is nothing other than
the entropy exchange $S_e(\pi_Q, \ccc N)$, 
such that the last relation becomes
\ben\label{number_eps_n} 
N_{\eps,n} \: \le \: 2^{n( S_e(\pi_Q, \ccc N) + \eps)}\:.
\een
To see this, we notice that $H(\ccc A_{\ccc N})$ equals the von Neumann entropy
of an $N$-dimensional diagonal density matrix $W$ with elements 
$W_{ii} = \fr{1}{M} \tr A_i\dag A_i$.
Since we are working in a diagonal operator-sum representation, this
actually means that 
$ W_{ij} = \fr{1}{M} \tr A_j\dag A_i = \tr A_i \pi_Q A_j\dag $, 
where $\pi_Q = \b 1_Q /M$.
By Schumacher's representation of the entropy exchange we thus realize
that $ H(\ccc A_{\ccc N}) = S_e(\pi_Q, \ccc N)$. 
 
Despite its strongly reduced number of Kraus operators, 
in average the operation $N_{\eps,n}$ does not much reduce the trace
when  $n$ becomes large. 
This can be seen by the selection probability $\tr  \ccc
N_{\eps,n}(\pi_{Q_n})$ of the homogeneously distributed state 
$\pi_{Q_n} = \b 1_{Q_n} / M^n$. 
A lower bound can be derived by observing that 
\be
\tr \ccc N_{\eps,n} (\pi_{Q_n}) 
= 
\fr{1}{M^n} \sum_{ A_{\b j} \scrbox{ $\eps$-typical}} \tr A_{\b j} A_{\b j}\dag
=
\sum_{ A_{\b j} \scrbox{ $\eps$-typical}}  p_{\b j} \:
\ee
is the probability that an operation element $A_{\b j}$ of $\ccc
N\tenpow n$ is $\eps$-typical. Thus, by 
Sec.\ \ref{subsubsec-typical_sequences}, property 2., 
\ben\label{reduced_trace}
\tr\: \ccc N_{\eps,n}( \pi_{Q_n}) \: \ge \: 1 - 2 e^{-n \psi_1(\eps)}\:,
\een
where $\psi_1(\eps)$ is a positive number independent of $n$. 

\subsubsection{Projection on typical subspace}\label{subsec-projection}
We will further reduce the operation $\ccc N\tenpow n$ by 
letting $\ccc N_{\eps,n}$ follow a 
projection on the $\eps$-typical subspace  $T_{\eps,n} \subset
H_Q\tenpow n$ of the density $\ccc N(\pi_Q)$. The benefit of this
procedure is that the so obtained operation $\tilde{\ccc N}_{\eps,n}$
maps $\pi_{Q_n}$ to an almost homogeneously distributed state on
$T_{\eps,n}$, and thus establishes a situation similar to (ii) in
Sec.\ \Ref{sec-general-noise}. 

The $\eps$-typical subspace $T_{\eps,n} \subset H_Q\tenpow n$ of
$\sigma \equiv \ccc N(\pi_Q)$ is spanned by the $\eps$-typical
eigenvectors of $\sigma \tenpow n$ \cite{NC00}.
These are precisely the eigenvectors $v_{\b l}$
with eigenvalues $p_{\b l}$ satisfying 
\be
2^{-n( S(\ccc N(\pi_Q)) + \eps)} \: \le \: p_{\b l} \: \le \: 
2^{-n( S(\ccc N(\pi_Q)) - \eps)}\:.
\ee
The dimension of $T_{\eps,n}$ obeys
\ben\label{typical_dimension}
\dim T_{\eps,n} \le 2^{n( S( \ccc N(\pi_Q)) + \eps)}\:.
\een
If $n$ is large, almost the entire weight of $\sigma\tenpow n$ lies in
the $\eps$-typical subspace: Let $\Pi_{\eps,n}$ be the projection on
$T_{\eps,n}$, then 
\be
\tr\: \Pi_{\eps, n} \sigma\tenpow n = 
\sum_{ 
{\scriptsize \mbox{ $\b l$ : $\ket{ v_{\b l}}$ $\eps$-typical}     }
}   p_{\b l}\:, 
\ee
which in the notions of Sec.\ \ref{subsubsec-typical_sequences} is the 
probability that an eigenvalue $\ket{ v_{\b l}} = 
\ket{ v_{l_1} } \ket{ v_{l_2} } \dots \ket{ v_{l_M} }$ 
is $\eps$-typical. Thus, by the second property in
Sec.\ \ref{subsubsec-typical_sequences}, 
\ben\label{typical_trace}
\tr\: \Pi_{\eps,n} \sigma \tenpow n \: \ge \: 1 - 2 e^{- n \psi_2(\eps)}\:,
\een 
where $\psi_2(\eps)$ is a positive number independent of $n$.

We define the {\em $\eps$-reduced operation of $\ccc N\tenpow n$} by 
\be
\tilde{\ccc N}_{\eps, n} := \ccc P_{\eps,n} \circ \ccc N_{\eps,n}\:,
\ee
where the operation $\ccc P_{\eps,n}$ describes the projective
measurement on $T_{\eps,n}$, 
\be
\ccc P_{\eps,n} : \rho \mapsto \Pi_{\eps,n} \:\rho\:
\Pi_{\eps,n}\:,
\ee
and $\ccc N_{\eps,n}$ is the $\eps$-typical operation of $\ccc N\tenpow n$
as defined in the previous subsection. 

\subsubsection{Properties of the $\eps$-reduced operation $\tilde{\ccc
    N}_{\eps,n}$}\label{subsubsec-properties}  
The $\eps$-reduced operation $\tilde{\ccc N}_{\eps,n}$ can be
represented by Kraus operators of the form 
$ \Pi_{\eps,n} A_{\b j} $, where $A_{\b j}$ is an $\eps$-typical
operation element of $\ccc N\tenpow n$. Their total number $\tilde
N_{\eps,n}$ is therefore bounded by
\be 
\tilde N_{\eps,n} = N_{\eps,n} \: \le \: 2^{n( S_e(\pi_Q,\ccc N) + \eps)}\:.
\ee
Besides the number of Kraus operators, the two other crucial figures  
are $\tr \tilde{\ccc N}_{\eps,n}(\pi_{Q_n})$ and 
$ \fn{ \tilde{\ccc N}_{\eps, n}(\pi_{Q_n}) }^2 $
(cf.\ relation  \Ref{fe-tildeN-V}).
In Appendix \ref{app-properties} we derive the followings bounds:
\bea
\tr\: \tilde{\ccc N}_{\eps,n} ( \pi_{Q_n} )  
&\: \ge \: &
1 - 4 e^{-n \psi_3(\eps)}\:, \\
\fn{ \tilde{\ccc N}_{\eps,n} (\pi_{Q_n}) }^2  
&\: \le \: &
 2^{-n ( S( \ccc N(\pi_Q)) - 3 \eps)}\:, 
\eea
where $\psi_3(\eps)$ is a positive number independent of $n$.
Finally, we note that for any code $C \subset H_Q\tenpow n$ 
\be
F_e(C, \ccc N\tenpow n) 
\: \ge \: 
F_e(C, \ccc N_{\eps,n}) 
\: \ge \:
F_e(C, \tilde{\ccc N}_{\eps,n})\:.
\ee
The first inequality holds because $\ccc N_{\eps,n}$ is a reduction of
$\ccc N\tenpow n$ and the second one is explained by the fact that 
$\tilde{\ccc N}_{\eps,n}$ 
results from post-processing of $\ccc N_{\eps,n}$ by $P_{\eps,n}$, which
cannot increase the code entanglement fidelity
(cf.\ Eq.\ \Ref{code-entanglement-fidelity}).

\subsection{Lower bounds for $Q(\ccc N) $ }\label{subsec-lower_boundsQ}
Lower bounds of the quantum capacity $Q(\ccc N)$ are given by the
achievable rates 
of $\ccc N$. Finding out whether a rate $R$ is achievable or
not requires to investigate the code entanglement fidelities $F_e(C_n, \ccc
N\tenpow n)$ for suitable codes $C_n \subset H_Q\tenpow n$ 
(cf.\ Sec.\ \ref{subsec-quantum_capacity}). 
Our working hypothesis is that no special care has to be taken in 
choosing $C_n$. Rather, we suppose that randomly chosen codes in general do 
provide high achievable rates and therefore will study the averaged 
entanglement fidelity of the code ensembles introduced in
\ref{subsec-unitarily_invariant}. 

\subsubsection{$Q(\ccc N) \ge I(\pi_Q, \ccc N)$}\label{subsubsec-rhoQ}
We begin with the average code fidelity $\cav{ F_e(C_n, \ccc N\tenpow
  n)}_{U_K}$ of the unitarily 
invariant ensemble $U_{K_n}$. As in 
\ref{subsec-discussionII}, we chose the code dimension to be 
\be
K_n = \lfloor 2^{n R} \rfloor\:,
\ee
meaning that  $R = \lim_{n\to\infty} \fr{1}{n} \log_2 K_n $ is the
asymptotic rate. 
By relation \Ref{fe-tildeN-V} and the results of the previous
subsection we immediately find
\ben\label{the-inequality}
\cav{ F_e( C, \ccc N\tenpow n ) }_{U_{K_n}} 
 \ge \: 
\cav{ F_e( C, \tilde{\ccc N}_{\eps,n}) }_{U_{K_n}}   
\ge \:
1 - \alpha_n - \beta_n
\een
with coefficients
\be
\begin{array}{rclcl} 
\alpha_n 
&= & 1 - \:\tr \: \tilde{\ccc N}_{\eps,n}(\pi_{Q_n}) 
 &\le&
 4\:e^{-n \psi_3(\eps)}\:, \\
\beta_n 
&=& \sqrt{ K_n \tilde N_{\eps,n}} \fn{ \tilde{\ccc N}_{\eps,n}
      (\pi_{Q_n})}
 &\le& 
  2^{\fr{n}{2}\left( R \:+\: S_e(\pi_Q, \ccc N) - S(\ccc N(\pi_Q)) \:
     + \:4 \eps  \right)}\:.
\end{array}
\ee
Clearly, for all $\eps > 0$, the right-hand side of inequality
\Ref{the-inequality} converges to unity in the limit $n \to \infty$ if
the asymptotic rate $R$ obeys 
\be
 R + 4 \eps \: < \: S(\ccc N(\pi_Q)) - S_e(\pi_Q, \ccc N) \equiv I(\pi_Q, \ccc N)\:.
\ee
That is, all rates $R$ below $I(\pi_Q,\ccc N)$ are achievable and
therefore $I(\pi_Q,\ccc N)$ is a lower bound of the capacity $Q(\ccc
N)$. 

\subsubsection{$Q(\ccc N) \ge I(\pi_V, \ccc N)   $ }\label{subsubsec-rhoV}
Let $V$ be an arbitrary linear subspace of the system's Hilbert space $H_Q$,
and let $\pi_V = \Pi_V /\dim V$. In short, the coherent information
$I(\pi_V, \ccc N)$ can be 
established as a lower bound of $Q(\ccc N)$ in exactly the same way as
before $I(\pi_Q, \ccc N)$ if we consider instead of $\ccc N$ the
operation $\ccc L$ that is defined as the restriction of $\ccc N$ to
states $\rho_V$ on a reduced input Hilbert space $V \subset H_Q$.
For the sake of completeness, we briefly repeat the arguments. 

This starts with reducing $\ccc L\tenpow n$ to an $\eps-$typical 
$\ccc L_{\eps,n}$ as described in Sec.\ \ref{subsec-typical} :
The reduced input Hilbert space $V$ of $\ccc L$ entails that 
now the probability $p_i$ of a Kraus operator $A_i$ has to be defined
as 
\ben\label{new_prob}
p_i = \fr{1}{L} \tr\: \Pi_V A_i\dag A_i \Pi_V\:, 
\een
where $L = \dim V$, and $\Pi_V$ is the projection on $V$. Here it is
assumed that the operators $A_1,\dots, A_N$ are diagonal with respect
to $V$, i.e.~$\tr \Pi_V A_i\dag A_j \Pi_V = 0$ for $i\neq j$. 
Accordingly, the probability of a $A_{\b j} = A_{j_1} \otimes \dots
\otimes A_{j_n}$ is 
\be
p_{\b j} = \fr{1}{L^n} \tr\: \Pi_V\tenpow n A_{\b j}\dag A_{\b j}
\Pi_V\tenpow n \:
= \: p_{j_1} \dots p_{j_n}\:.
\ee
As before, $\ccc L_{\eps,n}$ is defined to consist only of the
$\eps$-typical $A_{\b j}$. Its number $L_{\eps,n}$ is bounded
by $2^{n (H + \eps)}$, with $H$ being the Shannon entropy of the normalized
probability distribution \Ref{new_prob}. Therefore, $H$ conincides with the 
von Neumann entropy of a diagonal density matrix $W$ with entries
\be 
W_{ij} = \fr{1}{L} \tr\:\Pi_V A_i\dag A_j \Pi_V = \tr\: A_j
\fr{\Pi_V}{L} A_i\dag\:. 
\ee
By Schumacher's representation of the entropy exchange we obtain $H=
S_e(\pi_V, \ccc N)$, where $\pi_V = \Pi_V/L$.

The next step is to further reduce $\ccc L_{\eps,n}$ to an operation
$\tilde{\ccc L}_{\eps,n}$ by projecting the output of $\ccc
L_{\eps,n}$ on the typical subspace $T_{\eps,n} \subset H_Q\tenpow n$
of the density $\ccc L(\pi_V) = \ccc N(\pi_V)$. This follows
precisely Sec.\ \ref{subsec-projection} with $\sigma= \ccc
N(\pi_Q)$ replaced by $\sigma = \ccc N(\pi_V)$. 
The resulting $\tilde{\ccc L}_{\eps,n}$ is characterized by (cf. Sec.\
\ref{subsubsec-properties}) 
\bea
\tilde L_{\eps,n} 
\: &\le & \: 
 2^{n(S_e(\pi_V, \:\ccc N) + \eps)}\:,  \\
\tr\: \tilde{ \ccc L}_{\eps,n} 
\: &\ge & \:
 1 - 4 e^{-n \psi_3(\eps)}, \\
\fn{ \tilde{ \ccc L}_{\eps,n}}^2 
\: &\le & \: 2^{-n( S(\ccc N(\pi_V)) - 3
  \eps)}\:. \\
F_e(C, \ccc L \tenpow n)
\: & \ge & \: F_e(C, \tilde{\ccc L}_{\eps,n} )\:,
\eea
where $\tilde L_{\eps,n}$ is the number of Kraus operators that is
needed to represent $\tilde{\ccc L}_{\eps,n}$.
Thus, by inequality \Ref{fe-tildeN-V}, 
\be
\cav{ F_e( C, \ccc L\tenpow n ) }_{U_{K_n}(V\tenpow n)} 
\: \ge \:
1 - \alpha_n - \beta_n\:,
\ee
where the coefficients $\alpha_n$ and $\beta_n$ are as in the previous
subsection, but with $\pi_Q$ replaced by $\pi_V$. 
Since further $ \cav{ F_e(C, \ccc N\tenpow n)}_{U_{K_n}(V\tenpow n)} = 
 \cav{ F_e(C, \ccc L\tenpow n)}_{U_{K_n}(V\tenpow n)}  $ 
we can thus conclude that all
rates $R$ below 
\be
 S(\ccc N(\pi_V)) - S_e(\pi_V, \ccc N) \equiv I(\pi_V, \ccc N)
\ee
are achievable by $\ccc N$, meaning that $Q(\ccc N) \ge I(\pi_V, \ccc N)$. 

\subsubsection{$Q(\ccc N) \ge I_r(\ccc N)  $ }\label{subsubsec-rho}
Finally, we will show that with the BSST lemma 
the result of the last subsection implies the lower bound
\be
Q(\ccc N) \ge \fr{1}{m} I(\rho, \ccc N\tenpow m)\:,
\ee
where $m$ is an arbitrary large integer, and $\rho$ any density on
$H_Q\tenpow m$. Clearly, this suffices to prove the regularized
coherent information $I_r(\ccc N)$ (cf.\ Sec.\ \ref{subsec-quantum_capacity} )
a lower bound of $Q(\ccc N)$.  

The BSST lemma \cite{BSST02} states that for a channel $\ccc N$ and an
arbitrary state $\rho$ on the input space of $\ccc N$ 
\be
\lim_{\eps \to 0} \lim_{n \to \infty} \fr{1}{n} S(\ccc N\tenpow n(
\pi_{\eps,n}) ) \: = \: S(\ccc N(\rho))\:, 
\ee
where $\pi_{\eps,n}$ is the homogeneously distributed state on the
frequency-typical subspace $T^{(f)}_{\eps,n}$ of $\rho$.  
As a corollary, one obtains an analogous relation for
the coherent information, 
\be
\lim_{\eps \to 0} \lim_{n \to \infty} \fr{1}{n} I(\pi_{\eps,n}, \ccc
N\tenpow n) 
 \: = \: I( \rho, \ccc N )\:. 
\ee
$T^{(f)}_{\eps,n}$ is similar to the ordinary typical
subspace $T_{\eps,n}$ which we have used above. The difference is
that for $T^{(f)}_{\eps,n}$ typicality of a sequence is defined via 
the relative frequency of symbols in this sequence, whereas for
$T_{\eps,n}$
it is defined by its total probability. 
For details we refer the reader to the work of Holevo \cite{Hol02},
where an elegant proof of the BSST lemma is given. 

Here, what matters is solely the fact that $\pi_{\eps,n}$ is a 
homogeneously distributed subspace density of the kind that we
used in the previous subsection.  
Thus we can make use of the bound $Q(\ccc E) \ge I(\pi_V, \ccc E)$ 
with, for instance, $\ccc E = \ccc N \tenpow {mn}$, and $V$ being   
the frequency-typical subspace $T_{\eps,n}^{(f)} \subset H_Q\tenpow
{mn}$ of an arbitrary density $\rho$ on $H_Q\tenpow m$. This means
that for any  
$\eps > 0$ and any $m,n$ 
\be
Q( \ccc N \tenpow {mn} ) \ge I( \pi_{\eps,n}, \ccc N \tenpow {mn})\:.
\ee
Using the trivial identity $Q(\ccc N\tenpow k) = k Q(\ccc N)$ we can
therefore write  
\bea
Q(\ccc N) &=& \fr{1}{m} \lim_{n\to\infty} \fr{1}{n} Q(\ccc
N\tenpow{mn})\\
& \ge  &
\fr{1}{m} \lim_{\eps \to 0} \lim_{n \to \infty} 
  \fr{1}{n} I( \pi_{\eps,n}, (\ccc N \tenpow m) \tenpow n) \\
& =& 
 \fr{1}{m} I(\rho, \ccc N\tenpow m)\:,
\eea
where the last equation follows from the corollary.

\section{Concluding remarks}\label{sec-summary}
We expect that the 
lower bound \Ref{final-Fe-inequality} 
for the code entanglement fidelity 
is also useful for directly evaluating the error correcting
capability of a particular code for a particular noise operation. 
In this case, there is no need to estimate the trace norm of the
operator $D$ by its Frobenius norm. The only reason why we used this
in general rather poor estimate here is that it enabled us 
to perform the ensemble average.

The above proof of the direct coding theorem shows that a randomly
chosen code of sufficiently large block-size is typically a {\em good}
quantum error correcting code.
Studying the properties of unitarily-invariant  
code ensembles might be therefore always a good thing to do 
when general aspects of QEC are of concern. \\

{\em Note added.} We would like to mention the recent eprint of Hayden
{\em et al.}\ \cite{HHYW07}, in which a similar proof of the
direct coding theorem has been independently obtained. \\

\subsection*{Acknowledments}
I am grateful to M.\ R.\ Zirnbauer for pointing out the
use of group invariants. I would also like to thank the referee for
suggesting an improved lower bound for the code entanglement fidelity.

\begin{appendix}

\section{Fidelity relations}\label{app-average-code-fidelity}

\subsection{$ F_e(\pi_C, \ccc E) \le F_{av}(C, \ccc
  E)$}\label{subapp-fe_le_fav}

The average fidelity of the code $C$ with respect to noise $\ccc E$ is 
defined as 
\be 
F_{av}(C, \ccc E) = \int_{\b U(C)} d \mu_C(U)\: F_{ch}( U \psi_0 U\dag, \ccc
E)\:,
\ee
where $F_{ch}(\rho, \ccc E) = F(\rho, \ccc E(\rho))$, $\psi_0$ is an
arbitrary pure state in $C$, and $\mu_C$ is the normalized Haar
measure on the group $\b U(C)$ of unitaries on the code space  $C$. 
For a complete ensemble $\psi_1, \dots, \psi_K$ of orthogonal pure states in
$C$, $K = \dim C $, we find
\bea
F_{av}(C, \ccc E)  
&=&  
\int_{\b U(C)} d \mu_C(U) \fr{1}{K} \: \sum_{i=1}^K F_{ch}( U \psi_i
U\dag, \ccc E) \\
& \ge &
\int_{\b U(C)} d \mu_C(U) \:  F_e \left(\fr{1}{K}\sum_{i=1}^K U \psi_i
U\dag, \ccc E \right) = F_e(\pi_C, \ccc E)\:.
\eea
The inequality follows from the general  relation \cite{NC00}
\ben\label{fe_le_fav}
\sum_i p_i F_{ch}(\rho_i, \ccc E) \: \ge \: F_e\left(\sum_i p_i \rho_i,
  \ccc E\right)\:.
\een

\subsection{Subcodes with high minimum fidelity}\label{subsec-subcodes}
Let $C$ be a code of dimension $K$ with entanglement fidelity 
\be
 F_e(\pi_C, \ccc E) = 1- \eps\:.
\ee
We will show that there is a subcode $\tilde{ C}$ of $C$ of
dimension  $\tilde K = \lfloor K/2 \rfloor $ with minimum fidelity 
\be 
F_{min}(\tilde{ C}, \ccc E) := \min_{\ket{\psi}\in C} F_{ch}(\psi, \ccc E)
\: \ge 1- 2 \eps\:. 
\ee

To this end, we recursively define a sequence of subspaces
$C_0 \supset C_1 \supset \dots \supset C_{K-1}$, and 
a corresponding sequence of code vectors $\ket{\psi_0}, \ket{\psi_1}, \dots,
\ket{\psi_{K-1}}$ as follows: 
\bea
i=0: \qquad\qquad   C_0 &:= & C \\
  \ket{\psi_0} &:=& \mbox{ vector of minimal fidelity in $C_0$} \\
i>0: \qquad\qquad C_{i} &:=& C_{i-1} \cap \ket{\psi_{i-1}}^\perp \\
 \ket{\psi_i} &:=& \mbox{ vector of minimal fidelity in $C_i$}
\eea
By construction, $\dim C_i = K-i$, and $F_{min}(C_i, \ccc E) = F(\psi_i, \ccc
E(\psi_i)) \equiv F_i$. It is also clear that the minimum vectors
$\ket{\psi_0}, \ket{\psi_1}, \dots \ket{\psi_{K-1}}$ form an
orthonormal basis of $C$. Hence $\pi_C = \fr{1}{K}
\sum_{i=0}^{K-1} \psi_i$, and,  by relation \Ref{fe_le_fav},  
\be
1- \eps \le \fr{1}{K} \sum_{i=0}^K F_i\:.
\ee
For any $0 < t<K$ we therefore obtain
\be
1-\eps 
\:\le \: 
\fr{1}{K} \sum_{i=0}^{K-1-t} F_i \:+\: \fr{1}{K}\sum_{i=K-t}^{K-1} F_i 
\: \le \: 
\fr{K-t}{K} + \fr{t}{K} F_{K-t}\:,
\ee
where the last inequality follows from
$1\ge F_0 \ge F_1 \ge \dots \ge F_{K-1} \ge 0$ and
\be
1-\eps 
\:\le \: 
\fr{K-t}{K} + \fr{t}{K} F_{K-t}\:
\ee
is equivalent to
\be
1-\fr{K}{t} \eps \le F_{K-t}\:,
\ee
meaning that subspace $C_{K-t}$ of dimension $t$ has
minimum fidelity larger than $1- \eps K/t$. Setting $t= \lfloor K/2 \rfloor$
completes the proof.

\section{Average of $| \sp{\psi|P|\psi}|^2  $} \label{app-rmt}
We show that independent of the normalized vector $\ket \psi \in H_Q$
\ben\label{to_prove}
\cav{| \sp{\psi|P|\psi}|^2 }_{U_K} = \fr{K^2 + K}{M^2 + M}
\een
(notations as in Sec.\ \ref{subsec-unitarily_invariant}).
By definition,
\be 
\cav{| \sp{\psi|P|\psi}|^2 }_{U_K} = \int d\mu(U) \: | \sp{\psi| UP_0
  U\dag |\psi}|^2\:,
\ee
where the integral extends over $\b U(H_Q)$ and $P_0$ is the projection
on an arbitrarily chosen linear subspace $C_0 \subset H_Q$ of
dimension $K$. We extend $\ket \psi \equiv \ket{\psi_1}$ to an orthonormal basis 
$\ket{ \psi_1 }, \dots, \ket{ \psi_M }$ of $H_Q$, and chose
\be
C_0 := \span\{ \ket{ \psi_1 }, \dots, \ket{ \psi_K }\}\:.
\ee
Then
\be
\cav{| \sp{\psi|P|\psi}|^2 }_{U_K} = \sum_{i,j=1}^K \int d\mu(U) \: 
|U_{1i}|^2 |U_{1j}|^2 \:,
\ee
where $U_{ij} = \sp{ \psi_i | U | \psi_j}$. Making use of the unitary
invariance of $\mu$, this becomes
\be
 K \int d\mu(U)\: |U_{11}|^4 
\: + \: 
(K^2-K) \int d\mu(U) \: |U_{11}|^2 |U_{12}|^2\:.
\ee
For the calculation of these integrals we refer to the work of Pereyra
and Mello \cite{PM83}, in which, amongst others, the joint probability
density for the elements  
$U_{11}, \dots, U_{1k}$ of a random unitary matrix $U\in U_K$ has been
determined to be  
\be 
p(U_{11}, \dots, U_{1k}) = c 
\left(1 - \sum_{a=1}^k |U_{1a}|^2 \right)^{n-k-1} 
\Theta(1 - \sum_{a=1}^k |U_{1a}|^2) \:, 
\ee
where $c$ is a normalization constant, and $\Theta(x)$ denotes the
standard unit step function. 
By a straightforward calculation, we obtain from this 
\bea
\int d\mu(U)\: |U_{11}|^4  &=& \fr{2}{M^2 + M}\:, \\
\int d\mu(U) \: |U_{11}|^2 |U_{12}|^2 &=& \fr{1}{M^2 + M}\:,
\eea
which immediately leads to Eq.\ \Ref{to_prove}.

\section{Lower bound for code
 entanglement fidelity}\label{app-selective-fidelity} 
Without loss of generality we can describe a possibly trace-decreasing $\ccc N$
as a unitary 
operation $U_{QE}$ on $QE$ which is followed by a projective measurement on $E$
that may reduce the trace. That is, for a general state $\rho_Q$
\be
\ccc N(\rho_Q) =  \tr_E\: (\b 1_Q \otimes P_W) U_{QE} \: \rho_Q
\otimes \psi_E \: U_{QE}\dag \:,
\ee 
where $\psi_E$ is a fixed initial pure state of $E$, and $P_W$
projects on some subspace of $H_E$. Let again $\psi_{RQ}$ be a
purification of $\rho_Q$, $\rho_R = \tr_Q \psi_{RQ}$, and 
let a normalized pure state $\psi_{RQE}'$ on $RQE$ be defined by 
its state vector
\be
\ket{\psi_{RQE}'} = \fr{1}{\sqrt{p}} 
(\b 1_{RQ} \otimes P_W) (\b 1_R \otimes U_{QE}) 
\:\ket{ \psi_{RQ}} \otimes  \ket{\psi_E}\:, 
\ee
where $p= \tr \ccc N(\rho_Q)$. The state $\psi_{RQE}'$ is purification 
of its properly normalized partial states $\rho_Q'$, $\rho_E'$,
$\rho_{RQ}'$, and $\rho_{RE}'$. Note that $\ccc N(\rho_Q) = p \rho_{RE}'$.  

Precisely as in Sec.~\Ref{sec-S-I} it follows that there exists a
recovery operation $\ccc R$ on $Q$ satisfying
\be 
F(\psi_{RQ}, \ccc I_R \otimes \ccc R(\rho_{RQ}')\:) \ge 1 - \tn{
  \rho_{RE}' - \rho_R  \otimes \rho_E'}\:.
\ee
By definition \Ref{selective-entanglement-fidelity} of entanglement
fidelity for trace-decreasing operations this immediately leads to
\be
F_e(\rho_Q, \ccc R \circ \ccc N) \: \ge \: p \: - 
\tn{
  p \rho_{RE}' - p \rho_R  \otimes \rho_E'}\:,
\ee
which generalizes relation \Ref{decoupling}.

Continuing in a similar manner as before in Sec.~\ref{sec-S-I}, we
consider $\rho_Q= \pi_C$ with the purification \Ref{psi_RQ}, 
and chose the unitary $U_{QE}$ with projection
$P_W$ such that  
\ben\label{PV_unitary_UQE}
(\b 1_Q \otimes P_W) U_{QE} \ket{ \psi_Q } \ket{1} = \sum_{i=1}^N A_i \ket{ \psi_Q}
\ket{i}\:,
\een
where $\ket 1 \equiv \ket{\psi_E}, \ket{2}, \dots, \ket{N}$ are
again orthonormal vectors in  $H_E$. Then, it is readily verified that
\bea
p\:\rho_{RE}' &=& \fr{1}{K} \sum_{ij=1}^N \sum_{l,m=1}^K
\tr_Q( A_i \ket{c_l^Q}\bra{c_m^Q} A_j\dag)\:
\ket{c_l^R}\bra{c_m^R}\otimes \ket{i}\bra{j}\:, \\
p\:\rho_R \otimes \rho_E' &=& \sum_{ij=1}^N \tr_Q( A_i \pi_C A_j\dag)\:
\rho_R \otimes \ket{i}\bra{j}\:,
\eea
where $p= \tr \ccc N(\pi_C)$, which precisely correspond to expressions \Ref{rho_RE'},
\Ref{rho_R'_times_rho_E'}.
As in Sec.~\Ref{sec-S-I} we conclude that 
\be
F_e(\pi_C, \ccc R \circ \ccc N) \: \ge \:  p \: - \tn{D}\:, 
\ee
showing that $\tr \ccc N(\pi_C) \: - \tn{D}$ is indeed a lower bound
of  $F_e(C, \ccc N)$. 

\section{Typical sequences}\label{app-typical}

The first property follows from 
\be 
1  =  
\sum_{\mbox{$\b A$}} p_{\b A} 
\: \ge \:
\sum_{\mbox{$\b A$ $\eps$-typical}} p_{\b A} 
\: \ge \:
N_{\eps,n} 2^{-n(H(\ccc A) +\eps)}\:.
\ee
To prove the second property we first realize that by definition
\bea
P_{\eps,n} 
&=&\Pr( \: \mbox{``$A_{j_1},\dots, A_{j_n}$ is $\eps$-typical''} \:) 
\: = \: \Pr( \left|-\log_2(p_{j_1}\dots p_{j_n}) - n H(\ccc A) \right| \:
\le\: n \eps) \\
&=& \Pr(\: | \sum_{l=1}^n \left(-\log_2 p_{j_l} - H(\ccc A)\right) |\: 
\le\: n \eps\:)\:. 
\eea
The negative logarithms of the probabilities $p_{j_l}$ can be
understood as $n$ independent random variables $Y_l$
that assume values $-\log_2 p_1, \dots, -\log_2 p_N$ with probabilities
$p_1, \dots, p_N$. Their mean is the Shannon entropy $H(\ccc A)$, 
\be
\mu = E(Y_1) = - \sum_{i=1}^{N} p_i \log_2 p_i = H(\ccc A)\:.
\ee
This means that 
\be
1 - P_{\eps,n}  
\: = \: \Pr( \: |\sum_{l=1}^n (Y_l - \mu)| \: \ge \: n \eps \:)
\ee
is the probability of a large deviation $\propto n$.
Since the variance $\sigma$ and all higher
moments of $Y_1-\mu$ are finite we can employ a result from the theory 
of large deviations \cite{GS92}, according to which 
\be
\Pr(\: |\sum_{l=1}^n (Y_l - \mu)| \: \ge \: n \eps \:)
\: \le  \: 2 e^{ -n \psi(\eps)}\:, 
\ee
where $\psi(\eps)$ is a positive number that is
approximately $\eps^2 / 2 \sigma^2$.

\section{Bounds for 
 $\tr \tilde{\ccc   N}_{\eps,n}(\pi_{Q_n}) $ 
 and 
 $\fn{ \tilde{\ccc   N}_{\eps,n}(\pi_{Q_n}) }^2$ 
}\label{app-properties}  
It is convenient to
introduce the complementary operation $\ccc M_{\eps,n}$ of $\c
N_{\eps,n}$ by 
\be 
\ccc N\tenpow n = \ccc N_{\eps,n} + \ccc M_{\eps,n}\:.
\ee 
The operation elements of $\ccc M_{\eps,n}$ are exactly the
$\eps$-``untypical'' operation elements of $\ccc N \tenpow n$.
Then, 
\bean
\tr\: \tilde{\ccc N}_{\eps,n} (\pi_{Q_n} ) 
&=&
       \tr\: \Pi_{\eps,n}( \ccc N \tenpow n (\pi_{Q_n})-\ccc M_{\eps,n}
       (\pi_{Q_n})) \nn\\ 
& \ge&  \tr  \Pi_{\eps,n} \ccc N \tenpow n (\pi_{Q_n}) 
     - \tr\: \ccc M_{\eps,n} (\pi_{Q_n})\:. \label{tr-tildeN-inequality}
\eean
The inequality results from the fact that for two positive operators 
$A,B$ always $\tr AB \ge 0 $, and therefore (indices suppressed)
\be
\tr\: \ccc M(\rho)= \tr\: \Pi \ccc M(\rho) + \tr\: (\b 1 - \Pi) \ccc M(\rho) \ge
\tr\: \Pi \ccc M(\rho)\:. 
\ee
The first term in Eq.\ \Ref{tr-tildeN-inequality} can be bounded from below as
\be
\tr\: \Pi_{\eps,n} \ccc N \tenpow n (\pi_{Q_n}) 
 = 
\tr\: \Pi_{\eps,n} \ccc N \tenpow n(\pi_Q \tenpow n)  
 = 
\tr\: \Pi_{\eps,n} (\ccc N(\pi_Q) )\tenpow n 
\: \ge \:
1 - 2 e^{-n \psi_2(\eps)}\:,
\ee
where we used inequality \Ref{typical_trace}. 
The second term in Eq.\ \Ref{tr-tildeN-inequality} obeys
\be
\tr\: \ccc M_{\eps,n}( \pi_{Q_n} )
\: = \:
\tr \: \ccc N \tenpow n ( \pi_{Q_n}) \: - \: \tr \: \ccc N_{\eps,n}(
\pi_{Q_n}) \: \le \: 2 e^{ -n \psi_1(\eps)}\:, 
\ee
by inequality \Ref{reduced_trace}. We thus find
\be
\tr\: \tilde{\ccc N}_{\eps,n} (\pi_{Q_n} )  
\: \ge \: 
1 - 2 (e^{-n \psi_2(\eps)} + e^{ -n \psi_1(\eps)}) 
\: \ge \:
1 - 4\: e^{-n \psi_3(\eps)}
\:,
\ee
when $\psi_3(\eps) := \min\{ \psi_1(\eps), \psi_2(\eps) \}$.
For large $n$ the homogeneously distributed state $\pi_{Q_n}$ is 
almost certainly selected by the reduced operation $\tilde{\ccc N}_{\eps,n}$.

Now, let us address the Frobenius norm of $\tilde{\ccc N}(\pi_{Q_n})$. 
For positive operators $A,B$
\be
\fn{ A + B}^2  
\: =   \: \fn{A}^2 + \fn{B}^2 + 2 \tr\: A B
\: \ge \: \fn{A}^2 + \fn{B}^2\:. 
\ee
This can be used to derive 
\be
\fn{ \ccc P_{\eps,n} \circ \ccc N \tenpow n (\pi_{Q_n}) }^2 
 = 
\fn{ \ccc P_{\eps,n} \circ ( \ccc N_{\eps,n} + \ccc
  M_{\eps,n})(\pi_{Q_n})}^2 
 \ge 
\fn{ \ccc P_{\eps,n} \circ \ccc N_{\eps,n}(\pi_{Q_n})}^2\:. 
\ee
Thus
\bea
\fn{ \tilde{\ccc N}_{\eps,n} (\pi_{Q_n}) }^2 
& = &
\fn{ \ccc P_{\eps,n} \circ \ccc N_{\eps,n}(\pi_{Q_n})}^2 \\
& \le &
\fn{  \ccc P_{\eps,n} \circ \ccc N \tenpow n (\pi_{Q_n}) }^2 \\
& = &
\fn{ \Pi_{\eps,n} \: ( \ccc N(\pi_Q) )\tenpow n \: \Pi_{\eps,n} }^2 \\
& = &
\sum_{\scrbox{ $\b l$ : $ \ket{v_{\b l}}$ $\eps$-typical } }
\left( p_{\b l} \right)^2 \\
&\le & 2^{-n ( S( \ccc N(\pi_Q)) - 3 \eps)}\:, 
\eea
where we used \Ref{typical_dimension} and $ p_{\b l} \le 2^{-n( S( \ccc N(\pi_Q)) -
\eps)}$ to derive the last inequality. 

\end{appendix}


\begin{thebibliography}{10}
\bibitem{jensens_inequality}
  If $h$ is a concave function, and $X$ a random
  variable in the domain of $h$, then by Jensen's inequality
  $
  \cav{ h(X) }  < h( \cav{X} )\:.
  $
\end{thebibliography}
\end{document}